\documentclass[smallcondensed]{svjour3}

\journalname{Journal of Automated Reasoning}

\usepackage[latin1]{inputenc}
\usepackage{mathptmx}
\usepackage{graphicx}
\usepackage{latexsym}
\usepackage{amssymb}
\usepackage{url}
\usepackage{mymacros}
\usepackage{infrules}

\newcommand{\cminor}{Cminor}
\newcommand{\Csharpminor}{C\#minor}
\newcommand{\compcert}{CompCert}

\def\eg{e.g.\ }

\def\etal{{\em et al.}\ }

\def\seq#1{#1^*}
\def\opt#1{#1^?}
\def\some#1{\lfloor #1 \rfloor}
\def\Clight{Clight{}}
\def\Csharpminor{C\#minor}
\def\tr{t} 
\def\tri{T}  
\def\beh{B}

\def\ls{\it sw}
\def\ofs{\delta} 
\def\op{{\it op}}
\def\chunk{\kappa} 

\def\fn{{\it Fd}}

\def\evv{{v_\nu}}
\def\ev{{\nu}}
\def\tr{{\it t}}
\def\E0{{\epsilon}}
\def\fl{\phi} 

\def\loc{\ell} 

\def\evallvalue{\Leftarrow} 
\def\evalexpr{\Rightarrow}
\def\evalstmt{\Rightarrow}
\def\evalstmtl{\Rightarrow}
\def\evalfunc{\Rightarrow}
\def\evalprog{\Rightarrow}
\def\outloop{\stackrel{loop}{\leadsto}}

\def\accessmode{{\cal A}}

\newcommand{\sevalexpr}[3]{#1, #2 \evalexpr #3}
\newcommand{\sevalexpl}[3]{#1, #2 \evallvalue #3}
\newcommand{\sexecstmt}[5]{#1, #2 \stackrel{#4}{\evalstmt} #3, #5}
\newcommand{\sexecstmtl}[5]{#1, #2 \stackrel{#4}{\evalstmtl} #3, #5}
\newcommand{\sevalfunc}[6]{#1(#2), #3 \stackrel{#4}{\evalfunc} #5, #6}
\newcommand{\sevalprog}[2]{ #1 \evalprog #2}
\newcommand{\sexecinfstmt}[3]{#1, #2 \stackrel{#3}{\evalstmt} \infty}
\newcommand{\sexecinfstmtl}[3]{#1, #2 \stackrel{#3}{\evalstmtl} \infty}
\newcommand{\sevalinffunc}[4]{#1(#2), #3 \stackrel{#4}{\evalfunc} \infty}

\newcommand{\Econdition}[3]{#1 ~\hbox{\tt?}~ #2 ~\hbox{\tt:}~ #3}

\newcommand{\None}{\emptyset}

\newcommand{\id}{\ensuremath{\mathit{id}}}
\newcommand{\out}{\ensuremath{\mathit{out}}}

\newcommand{\cat}[2]{#1.#2} 

\newcommand{\decl}{{\it dcl}}
\newcommand{\ef}{{\it Fe}}
\newcommand{\fd}{{\it Fd}}

\newcommand{\semiopen}[2]{\mathopen[ #1, #2 \mathclose[ }

\begin{document}
\title{Mechanized semantics for the \Clight{} subset of the C language
\thanks{This work was supported by Agence Nationale de la Recherche, grant
number ANR-05-SSIA-0019.}
}
\author{Sandrine Blazy \and Xavier Leroy}
\institute{%
S. Blazy \at
ENSIIE, 1 square de la Résistance, 91025 Evry cedex, France
\\ \email{Sandrine.Blazy@ensiie.fr}
\and
X. Leroy \at
INRIA Paris-Rocquencourt, B.P. 105, 78153 Le Chesnay, France
\\ \email{Xavier.Leroy@inria.fr}}

\maketitle

\begin{abstract}
This article presents the formal semantics of a large subset of the C
language called \Clight{}. \Clight{} includes pointer arithmetic,
{\tt struct} and {\tt union} types, C loops and structured {\tt switch}
statements. \Clight{} is the source language of the \compcert{}
verified compiler. The formal semantics of \Clight{} is a big-step
operational semantics that observes both terminating and diverging
executions and produces traces of input/output events.  The formal
semantics of \Clight{} is mechanized using the Coq proof assistant. In
addition to the semantics of \Clight{}, this article describes its
integration in the \compcert{} verified compiler and several ways by
which the semantics was validated.

\keywords{The C programming language
\and Operational semantics
\and Mechanized semantics
\and Formal proof
\and The Coq proof assistant
}
\end{abstract}

\section{Introduction}

Formal semantics of programming languages---that is, the mathematical
specification of legal programs and their behaviors---play an
important role in several areas of computer science.  For advanced
programmers and compiler writers, formal semantics provide a more
precise alternative to the informal English descriptions that usually
pass as language standards.  In the context of formal methods such as
static analysis, model checking and program proof, formal
semantics are required to validate the abstract interpretations and
program logics (\eg axiomatic semantics) used to analyze and reason
about programs.  The verification of programming tools such as
compilers, type-checkers, static analyzers and program verifiers is
another area where formal semantics for the languages involved is a
prerequisite.  While formal semantics for realistic languages can be
defined on paper using ordinary mathematics
\cite{SML,gurevich:C,borger:C}, machine assistance such as the use of
proof assistants greatly facilitates their definition and uses.

For high-level programming languages such as Java and functional
languages, there exists a sizeable body of mechanized formalizations
and verifications of operational semantics, axiomatic semantics, and
programming tools such as compilers and bytecode verifiers. Despite
being more popular for writing systems software and embedded software,
lower-level languages such as C have attracted less interest: several
formal semantics for various subsets of C have been published, but 
only a few have been mechanized.

The present article reports on the definition of the formal semantics
of a large subset of the C language called \Clight{}.  \Clight{}
features most of the types and operators of~C, including pointer
arithmetic, pointers to functions, and {\tt struct} and {\tt union} types, as
well as all C control structures except {\tt goto}.  The semantics of
\Clight{} is mechanized using the Coq proof assistant
\cite{Coq,Bertot-Casteran-Coqart}.  It is presented as a big-step
operational semantics that observes both terminating and diverging
executions and produces traces of input/output events.  The \Clight{}
subset of C and its semantics are presented in
sections~\ref{sec:syntC} and~\ref{sec:semC}, respectively.

The work presented in this paper is part of an ongoing project called
\compcert{} that develops a realistic compiler for the C
language and formally verifies that it preserves the semantics of the
programs being compiled. A previous
paper~\cite{2006-Leroy-Blazy-Dargaye} reports on the development and
proof of semantic preservation in Coq of the front-end of this compiler:
a translator from \Clight{} to \cminor{}, a low-level, imperative
intermediate language.  The formal verification of the back-end of
this compiler, which generates moderately optimized PowerPC assembly
code from \cminor{} is described in~\cite{2008-Leroy-backend}.
Section~\ref{sec:using-clight} describes the integration of the
\Clight{} language and its semantics within the \compcert{} compiler
and its verification.

Formal semantics for realistic programming languages are large and
complicated.  This raises the question of validating these semantics:
how can we make sure that they correctly capture the expected
behaviors?  In section~\ref{sec:validation}, we argue that the
correctness proof of the \compcert{} compiler provides an indirect but
original way to validate the semantics of \Clight{}, and discuss other
approaches to the validation problem that we considered.

We finish this article by a discussion of related work in
section~\ref{sec:sota}, followed by future work and conclusions in
section~\ref{sec:concl}.

\paragraph{Availability}
The Coq development underlying this article can be consulted on-line
at \hbox{\url{http://compcert.inria.fr}}.  

\paragraph{Notations}  $\semiopen x y$ denotes the semi-open interval of
integers $\{ n \in \mathbb{Z} \mid x \le n < y\}$.  For functions
returning ``option'' types, $\some x$ (read: ``some $x$'') corresponds
to success with return value $x$, and~$\None$ (read: ``none'')
corresponds to failure.  In grammars, $\seq a$ denotes 0, 1 or several
occurrences of syntactic category $a$, and $\opt a$ denotes an
optional occurrence of syntactic category~$a$.

\section{Abstract syntax of \Clight{}}\label{sec:syntC}

\Clight{} is structured into expressions, statements and functions.
In the Coq formalization, the abstract syntax is presented as
inductive data types, therefore achieving a deep embedding of \Clight{}
into Coq.

\subsection{Types} \label{sec:types}


\begin{figure}

\begin{syntaxleft}
\syntaxclass{Signedness:}
{\it signedness} & ::= &  {\tt Signed} \alt {\tt Unsigned}
\syntaxclass{Integer sizes:}
{\it intsize} & ::= &  {\tt I8} \alt {\tt I16} \alt {\tt I32} 
\syntaxclass{Float sizes:}
{\it floatsize} & ::= &  {\tt F32} \alt {\tt F64}
\syntaxclass{Types:}
\tau & ::= &  {\tt int} ({\it intsize}, {\it signedness}) \\
     & \alt & {\tt float} ({\it floatsize})  \\
     & \alt & {\tt void} \\
     & \alt & {\tt array} (\tau, n) \\
     & \alt & {\tt pointer} (\tau) \\
     & \alt & {\tt function} (\seq \tau, \tau) \\
     &\alt & {\tt struct} (\id, \fl) \\
     &\alt & {\tt union} (\id, \fl) \\
     &\alt & {\tt comp{\char95}pointer} (\id) 
\syntaxclass{Field lists:}
\fl & ::= & \seq {(\id, \tau)}
\end{syntaxleft}

\caption{Abstract syntax of Clight types.}
\label{fig:syntax-types}

\end{figure}

The abstract syntax of \Clight{} types is given in 
figure~\ref{fig:syntax-types}.  Supported types include arithmetic types
(integers and floats in various sizes and signedness),
array types, pointer types (including pointers to functions), function 
types, as well as {\tt struct} and {\tt union} types.  Named types
are omitted: we assume that {\tt typedef} definitions have been expanded
away during parsing and type-checking.

The integral types fully specify the bit size of integers and floats, 
unlike the C types {\tt int}, {\tt long}, etc, whose sizes are left largely
unspecified in the C standard.  Typically, the parser maps {\tt int} and
{\tt long} to size {\tt I32}, {\tt float} to size {\tt F32}, and {\tt double} to size {\tt F64}.
Currently, 64-bit integers and extended-precision floats are not supported.

Array types carry the number $n$ of elements of the array, as a
compile-time constant.  Arrays with unknown sizes ($\tau${\tt []} in C)
are replaced by pointer types in function parameter lists.  Their
only other use in C is within {\tt extern} declarations of arrays, which
are not supported in \Clight{}.

Functions types specify the number and types of the function arguments and the
type of the function result. Variadic functions and unprototyped
functions (in the style of Ritchie's pre-standard C)
are not supported.

In C, {\tt struct} and {\tt union} types are named and compared by name. This 
enables the definition of recursive {\tt struct} types such as
{\tt struct\ s1\ {\char123}\ int\ n;\ struct\ *\ s1\ next;{\char125}}.
Recursion within such types must go through a pointer type. For instance, the 
following is not allowed in C:
{\tt struct\ s2\ {\char123}\ int\ n;\ struct\ s2\ next;{\char125}}.
To obviate the need to carry around a typing environment mapping
{\tt struct} and {\tt union} names to their definitions, \Clight{} {\tt struct}
and {\tt union} types are structural: they carry a local identifier
$\id$ and the list $\fl$ of their fields (names and types).  Bit-fields are
not supported.  These types are compared by structure, like all other
\Clight{} types.  In structural type systems, recursive types are
traditionally represented with a fixpoint operator $\mu \alpha.\tau$,
where $\alpha$ names the type $\mu \alpha.\tau$ within $\tau$.  
We adapt this idea to \Clight{}: within a {\tt struct} or {\tt union} type,
the type ${\tt comp{\char95}pointer}(\id)$ stands for a pointer type to the nearest 
enclosing {\tt struct}  or {\tt union} type named $\id$. For example, the 
structure {\tt s1} defined previously in C is expressed by
$$ {\tt struct}({\tt s1}, ({\tt n}, {\tt int}({\tt I32},{\tt signed})) ({\tt next}, {\tt comp{\char95}pointer}({\tt s1})))$$
Incorrect structures such as {\tt s2} above cannot be expressed at all,
 since {\tt comp{\char95}pointer} let us refer to a pointer to an enclosing {\tt struct}
 or {\tt union}, but not to the {\tt struct} or {\tt union} directly.

\Clight{} does not support any of the type qualifiers of C ({\tt const},
{\tt volatile}, {\tt restrict}).  These qualifiers are simply erased during
parsing.

The following operations over types are defined: ${\tt sizeof}(\tau)$
returns the storage size, in bytes, of type $\tau$, and
${\tt field{\char95}offset}(\id,\fl)$ returns the byte offset of the field named
$\id$ in a {\tt struct} whose field list is $\fl$, or $\None$ if $\id$
does not appear in $\fl$.  The Coq development gives concrete
definitions for these functions, compatible with the PowerPC ABI
\cite[chap.~3]{PPC-ABI}.
Typically, {\tt struct} fields are laid out consecutively and padding is
inserted so that each field is naturally aligned.  Here are the only
properties that a \Clight{} producer or user needs to rely on:
\begin{itemize}
\item Sizes are positive: ${\tt sizeof}(\tau) > 0$ for all types $\tau$.
\item Field offsets are within the range of allowed byte offsets for
  their enclosing {\tt struct}:
  if ${\tt field{\char95}offset}(\id,\fl) = \some \ofs$ and $\tau$ is the type
  associated with $\id$ in $\fl$, then
$$\semiopen{\ofs}{\ofs + {\tt sizeof}(\tau)} \subseteq
  \semiopen{0}{{\tt sizeof}({\tt struct\ }\id'~\fl)}.$$
\item Different fields correspond to disjoint byte ranges:
  if ${\tt field{\char95}offset}(\id_i,\fl) = \some {\ofs_i}$ and $\tau_i$ is the type
  associated with $\id_i$ in $\fl$ and $\id_1 \not= \id_2$, then
$$\semiopen{\ofs_1}{\ofs_1 + {\tt sizeof}(\tau_1)} \inter
  \semiopen{\ofs_2}{\ofs_2 + {\tt sizeof}(\tau_2)} = \emptyset.$$
\item When a {\tt struct} is a prefix of another {\tt struct}, fields shared
  between the two {\tt struct} have the same offsets:
if ${\tt field{\char95}offset}(\id,\fl) = \some \ofs$, then 
${\tt field{\char95}offset}(\id,\fl.\fl') = \some \ofs$ for all additional fields
  $\fl'$.
\end{itemize}

\subsection{Expressions} \label{s:syntax-expressions}


\begin{figure}

\begin{syntaxleft}
\syntaxclass{Expressions:}
a & ::=  & \id & variable identifier\\
  & \alt & n & integer constant\\
  & \alt & f & float constant \\
  & \alt & {\tt sizeof}(\tau) & size of a type \\
  & \alt & \op_1 ~ a & unary arithmetic operation\\
  & \alt & a_1 ~\op_2 ~ a_2  & binary arithmetic operation\\
  & \alt & \hbox{{\tt *}} a & pointer dereferencing \\
  & \alt & a \dot \id   & field access \\
  & \alt & \hbox{{\tt \&}} a & taking the address of \\
  & \alt & (\tau) a & type cast \\
  & \alt & \Econdition{a_1}{a_2}{a_3} & conditional expressions
\syntaxclass{Unary operators:}
\op_1 & ::= & \hbox{{\tt -}} \alt \hbox{{\tt {\char126}}} \alt \hbox{{\tt !}}
\syntaxclass{Binary operators:}
\op_2 & ::=  & \hbox{{\tt +}} \alt \hbox{{\tt -}} \alt \hbox{{\tt *}} \alt \hbox{{\tt /}} \alt \hbox{{\tt \%}}
               & arithmetic operators \\ 
      & \alt & \hbox{{\tt <<}} \alt \hbox{{\tt >>}} \alt 
               \hbox{{\tt \&}} \alt \hbox{{\tt {\char124}}} \alt \hbox{{\tt {\char94}}}
               & bitwise operators\\
 &  \alt  & \hbox{{\tt <}} \alt \hbox{{\tt <=}} \alt \hbox{{\tt >}} \alt \hbox{{\tt >=}}
               \alt \hbox{{\tt ==}} \alt \hbox{{\tt !=}}
               & relational operators
\end{syntaxleft}
\caption{Abstract syntax of Clight expressions}
\label{fig:syntax}
\end{figure}

The syntax of expressions is given in figure~\ref{fig:syntax}.  
All expressions and their sub-expressions are annotated by their
static types.  In the Coq formalization, expressions $a$ are therefore
pairs $(b, \tau)$ of a type $\tau$ and a term $b$ of an inductive
datatype determining the kind and arguments of the expression.  In
this paper, we omit the type annotations over expressions, but write
${\tt type}(a)$ for the type annotating the expression $a$.
The types carried by expressions are necessary to determine the
semantics of type-dependent operators such as overloaded arithmetic
operators. The following expressions can occur in left-value position:
$\id$, $\hbox{{\tt *}}a$, and $a \dot \id$.

Within expressions, only side-effect free operators of C are supported,
but not assignment operators ({\tt =}, {\tt +=}, {\tt ++}, etc) nor function
calls.  In \Clight{}, assignments and function calls are presented as
statements and cannot occur within expressions.  As a consequence, all
\Clight{} expressions always terminate and are pure: their evaluation
performs no side effects.  The first motivation for this design
decision is to ensure determinism of evaluation.  The C standard
leaves evaluation order within expressions partially unspecified.  If
expressions can contain side-effects, different evaluation orders can
lead to different results.  As demonstrated by Norrish \cite{Norrish:phd},
capturing exactly the amount of nondeterminism permitted by the C
standard complicates a formal semantics.

It is of course possible to commit on a particular evaluation order in
a formal semantics for C.  (Most C compiler choose a fixed evaluation
order, typically right-to-left.)  This is the approach we followed in
an earlier version of this work \cite{2006-Leroy-Blazy-Dargaye}.  Deterministic
side-effects within expressions can be accommodated relatively easily
with some styles of semantics (such as the big-step operational
semantics of \cite{2006-Leroy-Blazy-Dargaye}), but complicate or even prevent other forms
of semantics.  In particular, it is much easier to define axiomatic
semantics such as Hoare logic and separation logic if expressions are
terminating and pure: in this case, syntactic expressions can safely
be used as part of the logical assertions of the logic.  Likewise,
abstract interpretations and other forms of static analysis are much
simplified if expressions are pure.  Most static analysis and program
verification tools for C actually start by pulling assignments and
function calls out of expressions, and only then perform analyses over
pure expressions
\cite{ccured2,caduceus,cute,framac,saturn,Hardekopf}.


\begin{figure}

\begin{syntaxleft}
\syntaxclass{Statements:}
s & ::=  & {\tt skip} & empty statement \\
  & \alt & a_1 = a_2 &  assignment \\  
  & \alt & a_1 = a_2(\seq a) & function call \\
  & \alt & a(\seq a) & procedure call \\
  & \alt & s_1 ; s_2 & sequence \\
  & \alt & {\tt if}(a) ~ s_1 ~ {\tt else} ~ s_2 & conditional \\
  & \alt & {\tt switch}(a) ~\ls & multi-way branch \\
  & \alt & {\tt while}(a) ~ s & ``while'' loop \\
  & \alt & {\tt do} ~ s ~ {\tt while} (a) & ``do'' loop \\
  & \alt & {\tt for}(s_1, a_2, s_3)~s & ``for'' loop \\
  & \alt & {\tt break} & exit from the current loop \\
  & \alt & {\tt continue} & next iteration of the current loop\\
  & \alt & {\tt return}~\opt a & return from current function
\syntaxclass{Switch cases:}
\ls & ::=  & {\tt default:}~s & default case \\
     & \alt & {\tt case\ }n:~s; \ls & labeled case
\end{syntaxleft}

\caption{Abstract syntax of Clight statements.}
\label{fig:syntax2}
\end{figure}

Some forms of C expressions are omitted in the abstract syntax but can
be expressed as syntactic sugar:
$$\begin{array}{lr@{~~}c@{~~}l}
\mbox{array access:} &
   a_1{\tt [}a_2{\tt ]} & \syntequal & \mbox{{\tt *}}(a_1 + a_2) \\
\mbox{indirect field access:} &
   a \mbox{{\tt ->}} \id & \syntequal & \mbox{{\tt *}}(a.\id) \\
\mbox{sequential ``and'':} &
   a_1 ~\hbox{{\tt \&\&}}~ a_2 & \syntequal &
   \Econdition{a_1}{(\Econdition{a_2}{1}{0})}{0} \\
\mbox{sequential ``or'':} &
   a_1 ~\hbox{{\tt {\char124}{\char124}}}~ a_2 & \syntequal &
   \Econdition{a_1}{1}{(\Econdition{a_2}{1}{0})}
\end{array}$$

\subsection{Statements}

Figure~\ref{fig:syntax2} defines the syntax of \Clight{} statements.
All structured control statements of C (conditional, loops, Java-style
{\tt switch}, {\tt break}, {\tt continue} and {\tt return}) are supported, but not
unstructured statements such as {\tt goto} and unstructured {\tt switch} like
the infamous ``Duff's device'' \cite{Duff-device}.
As previously mentioned, assignment $a_1 = a_2$ of an r-value $a_2$ to
an l-value $a_1$, as well as function calls, are treated as statements.
For function calls, the result can either be assigned to an l-value or
discarded.

Blocks are omitted because block-scoped variables are not supported in
\Clight{}: variables are declared either with global scope at the
level of programs, or with function scope at the beginning of
functions.

The {\tt for} loop is written ${\tt for}(s_1, a_2, s_3)~s$, where $s_1$ is
executed once at the beginning of the loop, $a_2$ is the loop condition,
$s_3$ is executed at the end of each iteration, and $s$ is the loop body.
In C, $s_1$ and $s_3$ are expressions, which are evaluated for their
side effects.  In \Clight{}, since expressions are pure, we use
statements instead.  (However, the semantics requires that these
statements terminate normally, but not by e.g. {\tt break}.)

A {\tt switch} statement consists in an expression and a list of cases.  A
case is a statement labeled by an integer constant (${\tt case\ }n$) or by
the keyword {\tt default}.  Contrary to C, the default case is mandatory
in a \Clight{} {\tt switch} statement and must occur last.

\subsection{Functions and programs}


\begin{figure}

\begin{syntaxleft}
\syntaxclass{Variable declarations:}
\decl & ::= & \seq{(\tau~\id)}  & name and type
\syntaxclass{Internal function definitions:}
F & ::= & \tau~\id(\decl_1) \, \{ \, \decl_2; \, s \, \} &
($\decl_1$ = parameters, $\decl_2$ = local variables)
\syntaxclass{External function declarations:}
\ef & ::= & {\tt extern\ } \tau~\id(\decl)
\syntaxclass{Functions:}
\fd & ::= & F \alt \ef & internal or external
\syntaxclass{Programs:}
P & ::= & \decl; \seq\fd; {\tt main} = \id & global variables,
functions, entry point
\end{syntaxleft}

\caption{Abstract syntax of Clight functions and programs.}
\label{fig:syntax3}
\end{figure}

A \Clight{} program is composed of a list of declarations for global
variables (name and type), a list of functions (see figure~\ref{fig:syntax3})
and an identifier naming the entry point of the program (the {\tt main} function in
 C). The Coq formalization supports a rudimentary
form of initialization for global variables, where an initializer is a
sequence of integer or floating-point constants; we omit this feature
in this article.

Functions come in two flavors: internal or
external.  An internal function, written 
$\tau~\id(\decl_1) \, \{ \, \decl_2; \, s \, \}$,
is defined within the language.  $\tau$ is the return type, $\id$ the
name of the function, $\decl_1$ its parameters (names and types),
$\decl_2$ its local variables, and $s$ its body.  
External functions ${\tt extern\ } \tau~\id(\decl)$ are merely declared,
but not implemented.  They are intended to model ``system calls'',
whose result is provided by the operating system instead of being
computed by a piece of \Clight{} code.  

\section{Formal semantics for \Clight{}}\label{sec:semC}

We now formalize the dynamic semantics of \Clight{}, using natural
semantics, also  known as big-step operational semantics.
The natural semantics observe the final result of program execution
(divergence or termination), as well as a trace of the invocations of
external functions performed by the program.  The latter represents
the input/output behavior of the program.  Owing to the restriction
that expressions are pure (section~\ref{s:syntax-expressions}), the
dynamic semantics is deterministic.

The static semantics of \Clight{} (that is, its typing rules) has not
been formally specified yet.  The dynamic semantics is defined without
assuming that the program is well-typed, and in particular without
assuming that the type annotations over expressions are consistent.
If they are inconsistent, the dynamic semantics can be undefined (the
program goes wrong), or be defined but differ from what the C standard
prescribes.

\subsection{Evaluation judgements}


\begin{figure}

\begin{syntaxleft}
\syntaxclass{Block references:} b & \in & \mathbb{Z}
\syntaxclass{Memory locations:}
\loc & ::= & (b,\ofs) & byte offset $\ofs$ (a 32-bit integer) within block $b$
\syntaxclass{Values:}
v & ::= & {\tt int}(n) & integer value ($n$ is a 32-bit integer) \\
  & \alt & {\tt float}(f) & floating-point value ($f$ is a 64-bit float) \\
  & \alt & {\tt ptr}(\loc) & pointer value\\
  & \alt & {\tt undef} & undefined value
\syntaxclass{Statement outcomes:}
\out & ::= & {\tt Normal} & continue with next statement \\
    & \alt & {\tt Continue} & go to the next iteration of the current loop \\
    & \alt & {\tt Break} & exit from the current loop \\
    & \alt & {\tt Return}  & function exit \\
    & \alt & {\tt Return}(v)  & function exit, returning the value $v$
\syntaxclass{Global environments:}
G & ::=  & (id \mapsto b) & map from global variables to block references \\
  &      & \times (b \mapsto \fn) & and map from function references
                                    to function definitions
\syntaxclass{Local environments:}
E & ::= &  \id \mapsto b   & map from local variables to block references
\syntaxclass{Memory states:}
M & ::= &  b \mapsto (lo, hi, \ofs \mapsto v) &
  map from block references to bounds and contents
\syntaxclass{Memory quantities:}
\chunk & ::= & \hbox to 0pt{${\tt int8signed} \alt {\tt int8unsigned}$\hss} \\
       & \alt& \hbox to 0pt{${\tt int16signed} \alt {\tt int16unsigned}$\hss} \\
       &\alt&  \hbox to 0pt{${\tt int32} \alt  {\tt float32} \alt {\tt float64} $\hss}
\syntaxclass{I/O values:}
\evv & ::= & {\tt int}(n) \alt {\tt float} (f)
\syntaxclass{I/O events:}
\ev & ::= & \id( \, \evv^* \, \mapsto \evv) &
name of external function, argument values, result value
\syntaxclass{Traces:}
\tr & ::= & \E0 \alt \ev.\tr & finite traces (inductive) \\
\tri & ::= & \E0 \alt \ev.\tri & finite or infinite traces (coinductive)
\syntaxclass{Program behaviors:}
\beh & ::=  & {\tt terminates}(\tr, n) & termination with trace $\tr$ and
exit code $n$ \\
     & \alt & {\tt diverges}(\tri) & divergence with trace $\tri$
\end{syntaxleft}%
Operations over memory states: \\
\begin{tabular}{lp{8cm}}
${\tt alloc}(M, lo, hi) = (M', b)$ &
Allocate a fresh block of bounds $\semiopen{lo}{hi}$. 
\\
${\tt free}(M, b) = M'$ &
Free (invalidate) the block $b$.
\\
${\tt load}(\chunk, M, b, n) = \some v $ &
Read one or several consecutive bytes (as determined by $\chunk$) at
block $b$, offset $n$ in memory state $M$.  If successful return the
contents of these bytes as value $v$.
\\
${\tt store}(\chunk, M, b, n, v) = \some{M'} $ &
Store the value $v$ into one or several consecutive bytes (as
determined by $\chunk$) at offset $n$ in block $b$ of memory state
$M$.  If successful, return an updated memory state $M'$.
\end{tabular}

\medskip

\noindent Operations over global environments: \\
\begin{tabular}{ll}
${\tt funct}(G, b) = \some b$ &
Return the function definition $\fn$ corresponding to the block b, if any.
\\
${\tt symbol}(G, \id) = \some b$ &
Return the block $b$ corresponding to the global variable or function
name $\id$.
\\
${\tt globalenv}(P) = G $ &
Construct the global environment $G$ associated with the program $P$.
\\
${\tt initmem}(P) = M $ &
Construct the initial memory state $M$ for executing the program $P$.
\end{tabular}

\caption{Semantic elements: values, environments, memory states,
  statement outcomes, etc}
\label{fig:semantics-common}
\end{figure}

The semantics is defined by the 10 judgements (predicates) listed
below.  They use semantic quantities such as values, environments,
etc, that are summarized in figure~\ref{fig:semantics-common} and
explained later.
$$\begin{array}{rcll}
G, E & |- & \sevalexpl {a} {M} {\loc} & \mbox{(evaluation of expressions in l-value position)} \\
G, E & |- & \sevalexpr {a} {M} {v} & \mbox{(evaluation of expressions in r-value position)} \\
G, E & |- & \sevalexpr {\seq a} {M} {\seq v} & \mbox{(evaluation of lists of expressions)}\\
G, E & |- & \sexecstmt {s} {M} {\out} {\tr} {M'} & \mbox{(execution of
  statements, terminating case)} \\
G, E & |- & \sexecstmtl {\ls} {M} {\out} {\tr} {M'} & \mbox{(execution
  of the cases of a {\tt switch}, terminating case)} \\
G    & |- & \sevalfunc {\fd} {\seq v} {M} {\tr} {v} {M'} &
\mbox{(evaluation of function invocations, terminating case)} \\
G, E & |- & \sexecinfstmt {s} {M} {\tri} & \mbox{(execution of
  statements, diverging case)} \\
G, E & |- & \sexecinfstmtl {\ls} {M} {\tri} & \mbox{(execution
  of the cases of a {\tt switch}, diverging case)} \\
G    & |- & \sevalinffunc {\fd} {\seq v} {M} {\tri} &
\mbox{(evaluation of function invocations, diverging case)} \\
     & \vdash & \sevalprog {P} {\beh} &  \mbox{(execution of whole programs)}
\end{array}$$

Each judgement relates a syntactic element
to the result of executing this syntactic element.
For an expression in l-value position, the result is a location $\loc$:
a pair of a block identifier $b$ and a byte offset $\ofs$ within
this block.  For an expression in r-value position and for a function
application, the result is a value $v$: the discriminated union of
32-bit integers, 64-bit floating-point numbers, locations
(representing the value of pointers), and the special value {\tt undef}
representing the contents of uninitialized memory.  \Clight{} does not
support assignment between {\tt struct} or {\tt union}, nor passing a {\tt struct}
or {\tt union} by value to a function; therefore, {\tt struct} and {\tt union}
values need not be represented.  

Following Norrish \cite{Norrish:phd} and Huisman and Jacobs
\cite{Huisman-Jacobs-00}, the result associated with the execution of
a statement $s$ is an {\em outcome} $\out$ indicating how the execution
terminated: either normally by running to completion or prematurely
via a {\tt break}, {\tt continue} or {\tt return} statement.

Most judgements are parameterized by a global environment $G$, a local
environment $E$, and an initial memory state $M$.  Local environments
map function-scoped variables to references of memory blocks
containing the values of these variables.  (This indirection through
memory is needed to allow the {\tt \&} operator to take the address of a
variable.)  These blocks are allocated at function entry and freed at
function return (see rule~\ref{rule:26} in figure~\ref{fig:dynsemfun}).
Likewise, the global environment $G$ associates block references to
program-global variables and functions.  It also records the
definitions of functions.

The memory model used in our semantics is detailed in
\cite{2008-Leroy-Blazy-memory-model}.  Memory states $M$ are modeled as a
collection of blocks separated by construction and identified by
integers $b$.  Each block has lower and upper bounds $lo, hi$, fixed
at allocation time, and associates values to byte offsets $\ofs \in
\semiopen {lo} {hi}$.  The basic operations over memory states are {\tt alloc},
{\tt free}, {\tt load} and {\tt store}, as summarized in
figure~\ref{fig:semantics-common}.

Since \Clight{} expressions are pure, the memory state is not modified
during expression evaluation.  It is modified, however, during the
execution of statements and function calls.  The corresponding
judgements therefore return an updated memory state $M'$.  They also
produce a trace $\tr$ of the external functions (system calls) invoked
during execution.  Each such invocation is described by an input/output event
$\ev$ recording the name of the external function invoked, the
arguments provided by the program, and the result value provided by
the operating system.

In addition to terminating behaviors, the semantics also characterizes
divergence during the execution of a statement or of a function call.
The treatment of divergence follows the coinductive natural approach
of Leroy and Grall \cite{2007-Leroy-Grall}.  The result of a diverging
execution is the trace $\tri$ (possibly infinite) of input/output events performed.


\begin{figure}

\numberrules

Expressions in l-value position:
\begin{pannel}

\irule
        E(\id) = b \mbox{ or } 
        (\id \notin \Dom(E) \mbox{ and } {\tt symbol}(G,\id) = \some b)
        --------------------------------------------------
        G, E |- \sevalexpl {\id} {M}  {(b, 0)} 
\end \label{rule:1}

\irule
        G, E |- \sevalexpr {a} {M} { {\tt ptr} (\loc) }
        --------------------------------------------------
        G, E |- \sevalexpl {\hbox{{\tt *}}a} {M}  {\loc}
\end \label{rule:2}


\irule
        G, E |- \sevalexpl {a} {M} {(b,\ofs)} &
        {\tt type}(a) = {\tt struct} (\id', \fl) &
       {\tt field{\char95}offset} (\id, \fl) =  \some{\delta'}
        --------------------------------------------------
        G, E |- \sevalexpl {a \dot \id} {M} {(b,\ofs + \delta')}
\end \label{rule:4}

\irule
        G, E |- \sevalexpl {a} {M} {\loc} &
        {\tt type}(a) = {\tt union} (\id', \fl) 
        --------------------------------------------------
        G, E |- \sevalexpl {a \dot \id} {M} {\loc}
\end \label{rule:5}

\end{pannel}
Expressions in r-value position:
\begin{pannel}

\srule
       G, E |- \sevalexpr {n} {M} {{\tt int} (n)}
\end \label{rule:60}

\srule
       G, E |- \sevalexpr {f} {M} {{\tt float} (f)}
\end \label{rule:61}

\srule
       G, E |- \sevalexpr {{\tt sizeof}(\tau)} {M}  {{\tt int}({\tt sizeof}(\tau))}
\end \label{rule:62}

\irule
        G, E |- \sevalexpl {a} {M} {\loc} &
        {\tt loadval} ({\tt type}(a), M', \loc) = \some{v}
        --------------------------------------------------
        G, E |- \sevalexpr {a} {M}  {v}
\end \label{rule:6}

\irule
        G, E |- \sevalexpl {a} {M} {\loc}
        --------------------------------------------------
        G, E |- \sevalexpr {\hbox{{\tt \&}}a} {M} {{\tt ptr} (\loc)}
\end \label{rule:7}

\irule
        G, E |- \sevalexpr {a_1} {M} {v_1} &
        {\tt eval{\char95}unop} (\op_1, v_1, {\tt type}(a_1)) =  \some{v}
        --------------------------------------------------
        G, E |- \sevalexpr {\op_1 ~ a_1} {M} {v}
\end \label{rule:63}

\irule
        G, E |- \sevalexpr {a_1} {M} {v_1} &
        G, E |- \sevalexpr {a_2} {M_1} {v_2} &
        {\tt eval{\char95}binop} (\op_2, v_1, {\tt type}(a_1), v_2, {\tt type}(a_2)) =  \some{v}
        --------------------------------------------------
        G, E |- \sevalexpr {a_1 ~\op_2 ~ a_2} {M} {v}
\end \label{rule:8}

\irule
        G, E |- \sevalexpr {a_1} {M} {v_1} &
        {\tt is{\char95}true} (v_1, {\tt type}(a_1)) &
        G, E |- \sevalexpr {a_2} {M} {v_2}
        --------------------------------------------------
        G, E |- \sevalexpr {\Econdition{a_1}{a_2}{a_3}}
                           {M} {v_2}
\end \label{rule:67}

\irule
        G, E |- \sevalexpr {a_1} {M} {v_1} &
        {\tt is{\char95}false} (v_1, {\tt type}(a_1)) &
        G, E |- \sevalexpr {a_3} {M} {v_3}
        --------------------------------------------------
        G, E |- \sevalexpr {\Econdition{a_1}{a_2}{a_3}}
                           {M} {v_3}
\end \label{rule:64}





\irule
        G, E |- \sevalexpr {a} {M} {v_1} &
        {\tt cast} (v_1, {\tt type}(a), \tau) = \some{v}
        --------------------------------------------------
        G, E |- \sevalexpr {(\tau) a} {M} {v}
\end \label{rule:9}
\end{pannel}

\caption{Natural semantics for Clight expressions}
\label{fig:dynsem1}
\end{figure}

In the Coq specification, the judgements of the dynamic semantics
are encoded as mutually inductive predicates (for terminating
executions) and mutually coinductive predicates (for diverging executions).
Each defining case of
each predicate corresponds exactly to an inference rule in the
conventional, on-paper presentation of natural semantics.
We show most of the inference rules in figures~\ref{fig:dynsem1} 
to~\ref{fig:dynsem5}, and explain them in the remainder of this section.

\subsection{Evaluation of expressions}

\paragraph{Expressions in l-value position}
The first four rules of figure~\ref{fig:dynsem1} illustrate the evaluation of 
an expression in l-value position. A variable $\id$ evaluates to the 
location $(b, 0)$, where $b$ is the block associated with $\id$ in
the local environment $E$ or the global environment $G$ (rule~\ref{rule:1}).
If an expression $a$ evaluates (as an r-value) to a pointer value
${\tt ptr}(\loc)$, then the location of the dereferencing expression
$\hbox{{\tt *}}a$ is $\loc$ (rule~\ref{rule:2}).  

For field accesses $a \dot \id$, the location
$\loc = (b,\ofs)$ of $a$ is computed.  If $a$ has {\tt union}
type, this location is returned unchanged. (All fields of a {\tt union}
share the same position.)  If $a$ has {\tt struct} type, the offset of
field $\id$ is computed using the {\tt field{\char95}offset} function, then added
to~$\delta$.

\paragraph{From memory locations to values}
The evaluation of an l-value expression $a$ in r-value position
depends on the type of $a$ (rule~\ref{rule:6}). If $a$ has scalar
type, its value is loaded from memory at the location of $a$.  If $a$
has array type, its value is equal to its location.  Finally, some
types cannot be used in r-value position: this includes {\tt void} in C
and {\tt struct} and {\tt union} types in \Clight{} (because of the
restriction that structs and unions cannot be passed by value).  To
capture these three cases, figure~\ref{fig:access} defines the
function $\accessmode$ that maps \Clight{} types to {\em access
modes}, which can be one of: ``by value'', with a memory quantity $\chunk$
(an access loads a quantity $\chunk$ from the address of the l-value);
``by reference'' (an access simply returns the address of the l-value);
or ``by nothing'' (no access is allowed).  The {\tt loadval} and
{\tt storeval} functions, also defined in figure~\ref{fig:access},
exploit address modes to implement the correct semantics for
conversion of l-value to r-value ({\tt loadval}) and assignment to an
l-value ({\tt storeval}).


\begin{figure}
\begin{syntaxleft}
\syntaxclass{Access modes:}
\mu & ::= & {\tt By{\char95}value}(\chunk) &  access by value\\
    & \alt & {\tt By{\char95}reference} & access by reference\\
    & \alt &  {\tt By{\char95}nothing} & no access
\end{syntaxleft}%
Associating access modes to \Clight{} types:
$$\begin{array}{rcl@{\quad}rcl}
\accessmode({\tt int}({\tt I8},{\tt Signed})) & = & {\tt By{\char95}value}({\tt int8signed}) &
\accessmode({\tt array}(\_,\_)) & = & {\tt By{\char95}reference} \\
\accessmode({\tt int}({\tt I8},{\tt Unsigned})) & = & {\tt By{\char95}value}({\tt int8unsigned}) &
\accessmode({\tt function}(\_,\_)) & = & {\tt By{\char95}reference} \\
\accessmode({\tt int}({\tt I16},{\tt Signed})) & = & {\tt By{\char95}value}({\tt int16signed}) &
\accessmode({\tt struct})(\_,\_)) & = & {\tt By{\char95}nothing} \\
\accessmode({\tt int}({\tt I16},{\tt Unsigned})) & = & {\tt By{\char95}value}({\tt int16unsigned}) &
\accessmode({\tt union})(\_,\_)) & = & {\tt By{\char95}nothing} \\
\accessmode({\tt int}({\tt I32},\_)) & = & {\tt By{\char95}value}({\tt int32}) &
\accessmode({\tt void}) & = & {\tt By{\char95}nothing} \\
\accessmode({\tt pointer}(\_)) & = & {\tt By{\char95}value}({\tt int32})
\end{array}$$
Accessing or updating a value of type $\tau$ at location $(b,\ofs)$ in memory
state $M$:
$$\begin{array}{rcll}
{\tt loadval}(\tau, M, (b, \ofs)) & = &
  {\tt load}(\chunk, M, b, \ofs) &
  \mbox{if $\accessmode(\tau) = {\tt By{\char95}value}(\chunk)$} \\
{\tt loadval}(\tau, M, (b, \ofs)) & = &
  \some{(b, \ofs)} &
  \mbox{if $\accessmode(\tau) = {\tt By{\char95}reference}$} \\
{\tt loadval}(\tau, M, (b, \ofs)) & = & 
  \None &
  \mbox{if $\accessmode(\tau) = {\tt By{\char95}nothing}$} \\
{\tt storeval}(\tau, M, (b, \ofs), v) & = &
  {\tt store}(\chunk, M, b, \ofs, v) &
  \mbox{if $\accessmode(\tau) = {\tt By{\char95}value}(\chunk)$} \\
{\tt storeval}(\tau, M, (b, \ofs), v) & = & 
  \None &
  \mbox{otherwise}
\end{array}$$

\caption{Memory accesses.}
\label{fig:access}
\end{figure}

\paragraph{Expressions in r-value position}
Rules~\ref{rule:60} to~\ref{rule:9} of figure~\ref{fig:dynsem1}
illustrate the evaluation of an expression in r-value position.
Rule~\ref{rule:6} evaluates an l-value expression in an r-value
context. The expression is evaluated to its location $\loc$. From this
location, a value is deduced using the {\tt loadval} function described
above.  By rule~\ref{rule:7}, $\hbox{{\tt \&}}a$ evaluates to the pointer
value ${\tt ptr}(\loc)$ as soon as the l-value $a$ evaluates to the
location $\loc$.

Rules~\ref{rule:63} and ~\ref{rule:8} describe the evaluation of unary
and binary operations.  Taking binary operations as an example, the
two argument expressions are evaluated and their values $v_1, v_2$ are
combined using the the {\tt eval{\char95}binop} function, which takes as
additional arguments the types $\tau_1$ and $\tau_2$ of the arguments,
in order to resolve overloaded and type-dependent operators.  To give
the general flavor of {\tt eval{\char95}binop}, here are the cases corresponding
to binary addition:
$$
\begin{array}{lllll}
\tau_1 & \tau_2 & v_1 & v_2 & {\tt eval{\char95}binop}(\hbox{{\tt +}}, v_1, \tau_1,
v_2, \tau_2) \\
\hline
{\tt int}(\_) & {\tt int}(\_) & {\tt int}(n_1) & {\tt int}(n_2) &
   \some{{\tt int}(n_1+n_2)} \\
{\tt float}(\_) & {\tt float}(\_) & {\tt float}(f_1) & {\tt float}(f_2) &
   \some{{\tt float}(f_1+f_2)} \\
{\tt ptr}(\tau) & {\tt int}(\_) & {\tt ptr}(b,\ofs) & {\tt int}(n) & 
   \some{{\tt ptr}(b, \ofs + n \times {\tt sizeof}(\tau))} \\
{\tt int}(\_) & {\tt ptr}(\tau) & {\tt int}(n) & {\tt ptr}(b,\ofs) & 
   \some{{\tt ptr}(b, \ofs + n \times {\tt sizeof}(\tau))} \\
\multicolumn{4}{c}{\mbox{otherwise}} & \None
\end{array}
$$
The definition above rejects mixed arithmetic such as ``int${}+{}$float''
because the parser that generates \Clight{} abstract syntax
(described in section~\ref{s:parser}) never produces this:
it inserts explicit casts from integers to floats in this case.
However, it would be easy to add cases dealing with mixed arithmetic.
Likewise, the definition above adds two single precision floats using
double-precision addition, in violation of the ISO C standard.  Again,
it would be easy to recognize this case and perform a single-precision
addition.

Rules~\ref{rule:67} and~\ref{rule:64} define the evaluation of
conditional expressions $\Econdition{a_1}{a_2}{a_3}$.
The predicates {\tt is{\char95}true} and {\tt is{\char95}false} determine the truth value of
the value of $a_1$, depending on its type.  At a {\tt float} type,
${\tt float}(0.0)$ is false and any other {\tt float} value is true.
At an {\tt int} or {\tt ptr} type, ${\tt int}(0)$ is false and ${\tt int}(n)$ ($n \not=0$)
and ${\tt ptr}(\loc)$ values are true. (The null pointer is represented as
${\tt int}(0)$.) All other combinations of values
and types are neither true nor false, causing the semantics to go wrong.


\begin{figure}

\numberrules

\begin{pannel}

\srule  G, E |- \sexecstmt {{\tt skip}} {M} {{\tt Normal}} {\E0} {M}
\end
\label{rule:10}

\srule  G, E |- \sexecstmt {{\tt break}} {M} {{\tt Break}} {\E0} {M}
\end
\label{rule:13}

\\
\srule  G, E |- \sexecstmt {{\tt continue}} {M} {{\tt Continue}} {\E0} {M}
\end
\label{rule:14}

\srule
        G, E |- \sexecstmt {({\tt return} ~ \None)} {M} {{\tt Return}} {\E0} {M}
\end \label{rule:20}

\\
\irule
        G, E |- \sevalexpr {a} {M} {v}
        --------------------------------------------------
        G, E |- \sexecstmt {({\tt return} ~ \some a)} {M} {{\tt Return} (v)} {\E0} {M}
\end \label{rule:21}

\irule
        G, E |- \sevalexpl {a_1} {M} {\loc} &
        G, E |- \sevalexpr {a_2} {M} {v} &
        {\tt storeval}({\tt type}(a_1), M, \loc, v) = \some{M'}
        --------------------------------------------------
        G, E |-  \sexecstmt {(a_1 \hbox{{\tt \ =\ }} a_2)} {M} {{\tt Normal}} {\E0} {M'}
\end \label{rule:12}

\\
\irule
        G, E |- \sexecstmt {s_1} {M} {{\tt Normal}} {\tr_1} {M_1} &
        G, E |- \sexecstmt {s_2} {M_1} {\out} {\tr_2} {M_2}
        --------------------------------------------------
        G, E |- \sexecstmt {(s_1 ; s_2)} {M} {\out} {\cat {\tr_1} \tr_2} {M_2}
\end \label{rule:15}

\irule
        G, E |- \sexecstmt {s_1} {M} {\out} {\tr} {M'} &
        \out \not= {\tt Normal}
        --------------------------------------------------
        G, E |- \sexecstmt {(s_1; s_2)} {M} {\out} {\tr} {M'}
\end \label{rule:16}
\end{pannel}
\caption{Natural semantics for Clight statements (other than loops and {\tt switch} statements)}
\label{fig:dynsem2}
\end{figure}

Rule~\ref{rule:9} evaluates a cast expression
$(\tau)a$.  The expression $a$ is
evaluated, and its value is converted from its natural type ${\tt type}(a)$ to
the expected type $\tau$ using the partial function {\tt cast}.  This
function performs appropriate conversions, truncations and
sign-extensions between integers and floats.  We take a lax
interpretation of casts involving pointer types: if the source and
destination types are both either pointer types or 32-bit {\tt int} types,
any pointer or integer value can be converted between these types
without change of representation.  However, the {\tt cast} function fails
when converting between pointer types and {\tt float} or small integer
types, for example.

\subsection{Statements and function invocations, terminating case}


\begin{figure}

\numberrules
Outcome updates (at the end of a loop execution):
$$
{\tt Break} \outloop {\tt Normal} \qquad
{\tt Return} \outloop {\tt Return} \qquad
{\tt Return} (v) \outloop {\tt Return} (v)
$$

{\tt while} loops:
\begin{pannel}
\irule
        G, E |- \sevalexpr {a} {M} {v} & {\tt is{\char95}false}(v, {\tt type}(a)) 
        --------------------------------------------------
        G, E |- \sexecstmt {({\tt while} (a) ~ s)} {M} {{\tt Normal}} {\E0} {M}
\end \label{rule:17}

\irule
        G, E |- \sevalexpr {a} {M} {v} & 
        {\tt is{\char95}true} (v, {\tt type}(a)) \\
        G, E |- \sexecstmt {s} {M} {\out} {\tr} {M'} &
       \out \outloop \out'
        --------------------------------------------------
        G, E |- \sexecstmt {({\tt while} (a)~s)} {M} {\out'} {\tr} {M'}
\end \label{rule:18}

\irule
        G, E |- \sevalexpr {a} {M} {v} & 
        {\tt is{\char95}true} (v, {\tt type}(a)) \\
        G, E |- \sexecstmt {s} {M} {({\tt Normal} \mid {\tt Continue})} {\tr_1} {M_1} &
        G, E |- \sexecstmt {({\tt while} (a)~s)} {M_1} {\out'} {\tr_2} {M_2} 
        --------------------------------------------------
        G, E |- \sexecstmt {({\tt while} (a)~s)} {M} {\out'} {\cat {\tr_1} \tr_2} {M_2}
\end \label{rule:19}
\end{pannel}

{\tt for} loops:
\begin{pannel}
\irule
        s_1 \neq {\tt skip} &
        G, E |- \sexecstmt {s_1} {M} {{\tt Normal}} {\tr_1} {M_1}  &
        G, E |- \sexecstmt {({\tt for} ({\tt skip},a_2,s_3) ~s)} {M_1} {\out} {\tr_2} {M_2}
        --------------------------------------------------
        G, E |- \sexecstmt {({\tt for} (s_1,a_2,s_3) ~ s)} {M} {\out} {\cat {\tr_1} \tr_2} {M_2}
\end \label{rule:40}

\irule
        G, E |- \sevalexpr {a_2} {M} {v} & 
        {\tt is{\char95}false} (v,{\tt type}(a_2))
        --------------------------------------------------
        G, E |- \sexecstmt {({\tt for} ({\tt skip},a_2,s_3) ~s)} {M} {{\tt Normal}} {\E0} {M}
\end \label{rule:41}

\irule
        G, E |- \sevalexpr {a_2} {M} {v} & 
        {\tt is{\char95}true} (v, {\tt type}(a_2)) \\
        G, E |- \sexecstmt {s} {M} {\out_1} {\tr_1} {M_1} &
       \out_1 \outloop \out
        --------------------------------------------------
        G, E |- \sexecstmt {({\tt for} ({\tt skip},a_2,s_3) ~s)} {M} {\out} {\tr} {M_1}
\end \label{rule:42}

\irule
        G, E |- \sevalexpr {a_2} {M} {v} & 
        {\tt is{\char95}true} (v, {\tt type}(a_2)) \\
        G, E |- \sexecstmt {s} {M} {({\tt Normal} \mid {\tt Continue})} {\tr_1} {M_1} &
        G, E |- \sexecstmt {s_3} {M_1} {{\tt Normal}} {\tr_2} {M_2} \\
        G, E |- \sexecstmt {({\tt for} ({\tt skip},a_2,s_3)~s)} {M_2} {\out} {\tr_3} {M_3} 
        --------------------------------------------------
        G, E |- \sexecstmt {({\tt for} ({\tt skip},a_2,s_3) ~)} {M} {\out} {\cat {\cat {\tr_1} \tr_2} \tr_3} {M_3}
\end \label{rule:43}

\end{pannel}
\caption{Natural semantics for Clight loops}
\label{fig:dynsem3}
\end{figure}

The rules in figure~\ref{fig:dynsem2} define the execution of a statement that
is neither a loop nor a {\tt switch} statement.
The execution of a {\tt skip} statement yields the {\tt Normal}
outcome and the empty trace (rule~\ref{rule:10}).
Similarly, the execution of a {\tt break} (resp. {\tt continue}) statement 
yields the {\tt Break} (resp. {\tt Continue})
outcome and the empty trace (rules~\ref{rule:13} and~\ref{rule:14}).
Rules~\ref{rule:20}--\ref{rule:21} describe the execution of a 
{\tt return} statement.
The execution of a {\tt return} statement evaluates the argument of the 
{\tt return}, if any, and yields a {\tt Return}
outcome and the empty trace.
 
Rule~\ref{rule:12} executes an assignment statement. An assignment statement 
$a_1 \hbox{{\tt \ =\ }} a_2$ evaluates
the l-value $a_1$ to a location $\loc$ and the r-value 
$a_2$ to a value $v$, then stores $v$ at $\loc$ using the {\tt storeval}
function of figure~\ref{fig:access}, producing the final memory state $M'$.
We assume that the types of $a_1$ and $a_2$ are identical, therefore
no implicit cast is performed during assignment, unlike in C.
(The \Clight{} parser described in section~\ref{s:parser} inserts an
explicit cast on the r-value $a_2$ when necessary.)
Note that {\tt storeval} fails if $a_1$ has a {\tt struct} or {\tt union} type:
assignments between composite data types are not supported in \Clight{}.

The execution of a sequence of two statements starts with the
execution of the first statement, thus yielding an outcome that
determines whether the second statement must be executed or not
(rules~\ref{rule:15}~and~\ref{rule:16}).
The resulting trace is the concatenation of both traces originating from
both statement executions.


\begin{figure}

\numberrules

Function calls:
\begin{pannel}

\irule
        G, E |- \sevalexpr {a_{fun}} {M} {{\tt ptr} (b,0)} &
        G, E |- \sevalexpr {a_{args}} {M} {v_{args}} \\
        {\tt funct} (G,b) = \some{\fd} &
        {\tt type{\char95}of{\char95}fundef} (\fd) = {\tt type}(a_{fun})  &
        G |- \sevalfunc {\fd} {v_{args}} {M}  {\tr} {v_{res}} {M'}
        --------------------------------------------------
        G, E |-  \sexecstmt {a_{fun}(a_{args})} {M} {v_{res}} {\tr} {M'}
\end \label{rule:25}

\irule
        G, E |- \sevalexpl {a} {M} {\loc} &
        G, E |- \sevalexpr {a_{fun}} {M} {{\tt ptr} (b,0)} &
        G, E |- \sevalexpr {a_{args}} {M} {v_{args}} \\
        {\tt funct} (G,b) = \some{\fd} &
        {\tt type{\char95}of{\char95}fundef} (\fd) = {\tt type}(a_{fun})  &
        G |- \sevalfunc {\fn} {v_{args}} {M} {\tr}  {v_{res}} {M_1} \\
        {\tt storeval} ({\tt type}(a), M_1, {\tt ptr}(\loc), v_{res}) = \some{M_2}
        --------------------------------------------------
        G, E |-  \sexecstmt {a = a_{fun}(a_{args})} {M} {v_{res}} {\tr} {M_2}
\end \label{rule:24}

\end{pannel}
Compatibility between values, outcomes and return types:
$$
{\tt Normal}, {\tt void}  \mathrel{\#}  {\tt undef} \qquad
{\tt Return}, \, {\tt void} \mathrel{\#} {\tt undef} \qquad
{\tt Return} (v), \, \tau \mathrel{\#} v \mbox{ when } \tau~\not={\tt void}
$$

Function invocations:
\begin{pannel}

\irule
        F = \tau~\id(\decl_1) \, \{ \, \decl_2; \, s \, \} \\
        {\tt alloc{\char95}vars} (M, \decl_1 + \decl_2,E) = (M_1, \seq b) &
        {\tt bind{\char95}params} (E, M_1, \decl_1, v_{args}) = M_2 \\
        G, E |-  \sexecstmt {s} {M_2} {\out} {\tr} {M_3} &
        \out, \, \tau \, {\tt \#} \, v_{res}
        --------------------------------------------------
        G |- \sevalfunc {F} {v_{args}} {M} {\tr} {v_{res}} {{\tt free} (M_3, \seq b)}
\end \label{rule:26}
\\
\irule
        \ef = {\tt extern\ } \tau~\id(\decl) &
        \ev = \id(\, v_{args}, v_{res})
        --------------------------------------------------
        G |- \sevalfunc {\ef} {v_{args}} {M} {\ev} {v_{res}} {M}
\end \label{rule:27}

\end{pannel}
\caption{Natural semantics for function calls}
\label{fig:dynsemfun}
\end{figure}

The rules in figure~\ref{fig:dynsem3} define the execution of {\tt while}
and {\tt for} loops. (The rules describing the execution of {\tt dowhile}
loops resemble the rules for {\tt while} loops and are omitted in this
paper.) Once the condition of a {\tt while} loop is evaluated to a value
$v$, if $v$ is false, the execution of the loop terminates normally,
with an empty trace (rules~\ref{rule:17} and~\ref{rule:41}). If $v$ is true,
the loop body $s$ is executed, thus yielding an outcome $\out$ 
(rules~\ref{rule:18},~\ref{rule:19},~\ref{rule:42} and~\ref{rule:43}). If 
$\out$ is {\tt Normal} or {\tt Continue},
the whole loop is re-executed in the memory state modified by the
first execution of the body. In $s$, the execution of a {\tt continue}
statement interrupts the current execution of the loop body and
triggers the next iteration of $s$.  If $\out$ is {\tt Break}, the loop
terminates normally; if $\out$ is {\tt Return}, the loop terminates
prematurely with the same outcome (rules~\ref{rule:18}
and~\ref{rule:42}).  The $\outloop$ relation models this evolution of
outcomes after the premature end of the execution of a loop body.

Rules~\ref{rule:40}--\ref{rule:43} describe the execution of a
${\tt for}(s_1, a_2, s_3)~s$ loop.
Rule~\ref{rule:40} executes the initial statement $s_1$ of a {\tt for} loop,
which must terminate normally.  Then, the loop with an empty
initial statement is executed in a way similar to that of a {\tt while}
loop (rules~\ref{rule:41}--\ref{rule:43}).  If the body $s$ terminates
normally or by performing a {\tt continue}, the statement $s_3$ is
executed before re-executing the {\tt for} loop.  As in the case of $s_1$,
it must be the case that $s_3$ terminates normally.

We omit the rules for ${\tt switch}(a) ~\ls$ statements, which are
standard.  Based on the integer value of $a$, the appropriate case of
$\ls$ is selected, and the corresponding suffix of $\ls$ is executed
like a sequence, therefore implementing the ``fall-through'' behavior
of {\tt switch} cases.  A {\tt Break} outcome for one of the cases terminates
the {\tt switch} normally.

The rules of figure~\ref{fig:dynsemfun} define the execution of a call
statement $a_{fun} (a_{args})$ or 
$a = a_{fun}(a_{args})$.  The expression $a_{fun}$ is evaluated to a
function pointer ${\tt ptr}(b, 0)$, and the reference $b$ is resolved to the
corresponding function definition $\fd$ using the global environment $G$.
This function definition is then invoked on the values of the
arguments $a_{args}$ as per the judgment
$G |- \sevalfunc {\fd} {v_{args}} {M} {v_{res}} {\tr} {M'}$.
If needed, the returned value $v_{res}$ is then stored in the location
of the l-value $a$ (rules~\ref{rule:25} and~\ref{rule:24}).

The invocation of an internal \Clight{} function $F$
(rule~\ref{rule:26}) allocates the memory required 
for storing the formal parameters and the local variables of $F$,
using the {\tt alloc{\char95}vars} function.  This function allocates one block
for each variable $\id: \tau$, with lower bound 0 and upper bound
${\tt sizeof}(\tau)$,  using the {\tt alloc} primitive of the memory model.
These blocks initially contain {\tt undef} values.  Then, the
{\tt bind{\char95}params} function iterates the 
{\tt storeval} function in order to initialize formal parameters to the
values of the corresponding arguments. 


\begin{figure}

\numberrules

\begin{pannel}

\iruledouble
G, E |- \sexecinfstmt {s_1} {M} {\tri}
-------------------------------------------
G, E |- \sexecinfstmt {s_1; s_2} {M} {\tri}
\end \label{rule:100}

\iruledouble
G, E |- \sexecstmt {s_1} {M} {{\tt Normal}} {\tr} {M_1} &
G, E |- \sexecinfstmt {s_2} {M_1} {\tri}
--------------------------------------------------
G, E |- \sexecinfstmt {s_1; s_2} {M} {\cat{\tr}{\tri}}
\end \label{rule:101}

\iruledouble
G, E |- \sevalexpr {a} {M} {v} & {\tt is{\char95}true} (v, {\tt type}(a)) &
G, E |- \sexecinfstmt {s} {M} {\tri}
-------------------------------------------------
G, E |- \sexecinfstmt {({\tt while} (a)~s)} {M} {\tri}
\end \label{rule:102}

\iruledouble
G, E |- \sevalexpr {a} {M} {v} & {\tt is{\char95}true} (v, {\tt type}(a)) \\
G, E |- \sexecstmt {s} {M} {({\tt Normal} \mid {\tt Continue})} {\tr} {M_1} &
G, E |- \sexecinfstmt {({\tt while} (a)~s)} {M_1} {\tri} 
--------------------------------------------------
G, E |- \sexecinfstmt {({\tt while} (a)~s)} {M} {\cat {\tr} {\tri}}
\end \label{rule:103}

\iruledouble
        G, E |- \sevalexpr {a_{fun}} {M} {{\tt ptr} (b,0)} &
        G, E |- \sevalexpr {a_{args}} {M} {v_{args}} \\
        {\tt funct} (G,b) = \some{\fd} &
        {\tt type{\char95}of{\char95}fundef} (\fd) = {\tt type}(a_{fun}) &
        G |- \sevalinffunc {\fd} {v_{args}} {M} {\tri}
        --------------------------------------------------
        G, E |-  \sexecinfstmt {a_{fun}(a_{args})} {M} {\tri}
\end \label{rule:104}

\iruledouble
        F = \tau~\id(\decl_1) \, \{ \, \decl_2; \, s \, \} \\
        {\tt alloc{\char95}vars} (M, \decl_1 + \decl_2,E) = (M_1, \seq b) &
        {\tt bind{\char95}params} (E, M_1, \decl_1, v_{args}) = M_2 \\
        G, E |-  \sexecinfstmt {s} {M_2} {\tri}
        --------------------------------------------------
        G |- \sevalinffunc {F} {v_{args}} {M} {\tri}
\end \label{rule:105}

\end{pannel}

\caption{Natural semantics for divergence (selected rules)}
\label{fig:dynsemdiv}
\end{figure}

The body of $F$ is then executed, thus yielding an outcome (fourth
premise).  The return value of $F$ is computed from this outcome and
from the return type of $F$ (fifth premise): for a function returning
{\tt void}, the body must terminate by {\tt Normal} or {\tt Return} and the return
value is {\tt undef}; for other functions, the body must terminate by
${\tt Return}(v)$ and the return value is $v$.  Finally, the memory blocks
$\seq b$ that were allocated for the parameters and local variables
are freed before returning to the caller.

A call to an external function $\ef$ simply generates an input/output event
recorded in the trace resulting from that call (rule~\ref{rule:27}).

\subsection{Statements and function invocations, diverging case}

Figure~\ref{fig:dynsemdiv} shows some of the rules that model
divergence of statements and function invocations.  As denoted by the
double horizontal bars, these rules are to be interpreted {\em
coinductively}, as greatest fixpoints, instead of the standard
inductive interpretation (smallest fixpoints) used for the other rules
in this paper.  In other words, just like terminating executions
correspond to finite derivation trees, diverging executions correspond
to infinite derivation trees \cite{2007-Leroy-Grall}.  

A sequence $s_1;s_2$ diverges either if $s_1$ diverges, or if $s_1$
terminates normally and $s_2$ diverges (rules~\ref{rule:100}
and~\ref{rule:101}). Likewise, a loop diverges either if its body
diverges, or if it terminates normally or by {\tt continue} and the next
iteration of the loop diverges (rules~\ref{rule:102}
and~\ref{rule:103}).  A third case of divergence corresponds to an
invocation of a function whose body diverges (rules~\ref{rule:104}
and~\ref{rule:105}).  

\subsection{Program executions}


\begin{figure}
\numberrules
\begin{pannel}

\irule
        G = {\tt globalenv} (P) &
        M = {\tt initmem} (P) \\
        {\tt symbol} (G,{\tt main}(P))= \some{b} &
        {\tt funct} (G,b)= \some{f} &
        G |- \sevalfunc {f} {{\tt nil}} {M} {\tr} {{\tt int}(n)} {M'}
        --------------------------------------------------
         |-  \sevalprog {P} {{\tt terminates}(\tr, n)}
\end \label{rule:50}

\irule
        G = {\tt globalenv} (P) &
        M = {\tt initmem} (P) \\
        {\tt symbol} (G,{\tt main}(P))= \some{b} &
        {\tt funct} (G,b)= \some{f} &
        G |- \sevalinffunc {f} {{\tt nil}} {M} {\tri}
        --------------------------------------------------
         |-  \sevalprog {P} {{\tt diverges}(\tri)}
\end \label{rule:51}

\end{pannel}
\caption{Observable behaviors of programs}
\label{fig:dynsem5}
\end{figure}

Figure~\ref{fig:dynsem5} defines the execution of a program $P$ and
the determination of its observable behavior.  A global environment
and a memory state are computed for $P$, where each global variable is
mapped to a fresh memory block.  Then, the main function of $P$ is
resolved and applied to the empty list of arguments.  If this function
invocation terminates with trace $\tr$ and result
value ${\tt int}(n)$, the observed behavior of $P$ is
${\tt terminates}(\tr,n)$ (rule~\ref{rule:50}).
If the function invocation diverges with a possibly infinite trace
$\tri$, the observed behavior is ${\tt diverges}(T)$ (rule~\ref{rule:51}).

\section{Using \Clight{} in the CompCert compiler} \label{sec:using-clight}

In this section, we informally discuss how \Clight{} is used in the
CompCert verified compiler
\cite{2006-Leroy-compcert,2006-Leroy-Blazy-Dargaye,2008-Leroy-backend}.

\subsection{Producing \Clight{} abstract syntax} \label{s:parser}

Going from C concrete syntax to \Clight{} abstract syntax is not as
obvious as it may sound.  After an unsuccessful attempt at
developing a parser, type-checker and simplifier from scratch, we
elected to reuse the CIL library of Necula \etal \cite{cil}.  CIL
is written in OCaml and provides the following facilities:
\begin{enumerate}
\item A parser for ISO C99 (plus GCC and Microsoft extensions),
  producing a parse tree that is still partially ambiguous.
\item A type-checker and elaborator, producing a precise,
  type-annotated abstract syntax tree.
\item A simplifier that replaces many delicate constructs of C by
  simpler constructs.  For instance, function calls and assignments
  are pulled out of expressions and lifted to the statement level.
  Also, block-scoped variables are lifted to function scope or global
  scope.
\item A toolkit for static analyses and transformations performed
  over the simplified abstract syntax tree.
\end{enumerate}
While conceptually distinct, (2) and (3) are actually performed in a
single pass, avoiding the creation of the non-simplified abstract
syntax tree.  

Thomas Moniot and the authors developed (in OCaml) a simple translator that
produces \Clight{} abstract syntax from the output of CIL.  Much
information produced by CIL is simply erased, such as type attributes
and qualifiers.  {\tt struct} and {\tt union} types are converted from the
original named representation to the structural representation used by
\Clight{}.  String literals are turned into global, initialized arrays
of characters.  Finally, constructs of C that are unsupported in \Clight{}
are detected and meaningful diagnostics are produced.

The simplification pass of CIL sometimes goes too far for our needs.
In particular, the original CIL transforms all C loops into
{\tt while(1)\ {\char123}\ ...\ {\char125}} loops, sometimes inserting {\tt goto} statements to
implement the semantics of {\tt continue}.  Such CIL-inserted {\tt goto}
statements are problematic in \Clight{}.  We therefore patched CIL to
remove this simplification of C loops and natively support {\tt while},
{\tt do} and {\tt for} loops

CIL is an impressive but rather complex piece of code, and it has not
been formally verified.  One can legitimately wonder whether we can
trust CIL and our hand-written translator to preserve the
semantics of C programs.  Indeed, two bugs in this part of CompCert
were found during testing: one that we introduced when adding native
support for {\tt for} loops; another that is present in the unmodified CIL
version 1.3.6, but was corrected since then. 

We see two ways to
address this concern.  First, we developed a pretty-printer that
displays \Clight{} abstract syntax tree in readable, C concrete
syntax.  This printer makes it possible to conduct manual reviews of the
transformations performed by CIL.  Moreover, experiment shows that
re-parsing and re-transforming the simplified C syntax printed from
the \Clight{} abstract syntax tree reaches a fixed point in one
iteration most of the time.  This does not prove anything but
nonetheless instills some confidence in the approach.

A more radical way to establish trust in the CIL-based \Clight{}
producer would be to formally verify some of the simplifications
performed.  A prime candidate is the simplification of expressions,
which transforms C expressions into equivalent pairs of a statement
(performing all side effects of the expression) and a pure expression
(computing the final value).  Based on initial experiments
on a simple ``while'' language, the Coq verification of this
simplification appears difficult but feasible.  We leave this line of
work for future work.

\subsection{Compiling \Clight{}} \label{s:compiling-clight}

The CompCert C compiler is structured in two parts: a front-end
compiler translates \Clight{} to an intermediate language called
Cminor, without performing any optimizations; a back-end compiler
generates PowerPC assembly code from the Cminor intermediate
representation, performing good register allocation and a few
optimizations.  Both parts are composed of multiple
passes.  Each pass is proved to preserve semantics: if the input
program $P$ has observable behavior $B$, and the pass translates $P$
to $P'$ without reporting a compile-time error, then the output
program $P'$ has the same observable behavior $B$.  The proofs of
semantic preservation are conducted with the Coq proof assistant.  To
facilitate the proof, the compiler passes are written directly in the
specification language of Coq, as pure, recursive functions.
Executable Caml code for the compiler is then generated automatically
from the functional specifications by Coq's extraction facility.

The back-end part of CompCert is described in great detail in
\cite{2008-Leroy-backend}.  We now give an overview of the
front-end, starting with a high-level overview of Cminor, its target
intermediate language.  (Refer to \cite[section 4]{2008-Leroy-backend}
for detailed specifications of Cminor.)

Cminor is a low-level imperative language, structured like \Clight{}
into expressions, statements, and functions.  A first difference with
\Clight{} is that arithmetic operators are not overloaded and their
behavior is independent of the static types of their operands:
distinct operators are provided for integer arithmetic and
floating-point arithmetic.  Conversions between integers and floats
are explicit.  Likewise, address computations are explicit in Cminor,
as well as individual load and store operations.  For instance, the C
expression {\tt a[x]} where {\tt a} is a pointer to {\tt int} is expressed as
{\tt load(int32,\ a\ +}$_i${\tt \ x\ *}$_i${\tt \ 4)}, making explicit the memory quantity
being addressed ({\tt int32}) as well as the address computation.

At the level of statements, Cminor has only 5 control structures:
if-then-else conditionals, infinite loops, {\tt block}-{\tt exit}, early
return, and {\tt goto} with labeled statements.  The ${\tt exit}~n$ statement
terminates the $(n+1)$ enclosing {\tt block} statements.

Within Cminor functions, local variables can only hold scalar
values (integers, pointers, floats) and they do not reside in
memory.  This makes it easy to allocate them to registers later in the
back-end, but also prohibits taking a pointer to a local variable
like the C operator {\tt \&} does.  Instead, each Cminor function declares
the size of a stack-allocated block, allocated in memory at function
entry and automatically freed at function return.  The expression
${\tt addrstack}(n)$ returns a pointer within that block at constant
offset $n$.  The Cminor producer can use this block to store local
arrays as well as local scalar variables whose addresses need to be
taken.

To translate from \Clight{} to Cminor, the front-end of CompCert~C
therefore performs the following transformations:
\begin{enumerate}
\item Resolution of operator overloading and materialization of all
type-dependent behaviors.  Based on the types that annotate \Clight{}
expressions, the appropriate flavors (integer or float) of arithmetic
operators are chosen; conversions between ints and floats, truncations and
sign-extensions are introduced to reflect casts; address computations
are generated based on the types of array elements and pointer
targets; and appropriate memory chunks are selected for every memory
access.
\item Translation of {\tt while}, {\tt do} and {\tt for} loops into
infinite loops with blocks and early exits.  The {\tt break}
and {\tt continue} statements are translated as appropriate {\tt exit} constructs.
\item Placement of Clight variables, either as Cminor local variables
(for local scalar variables whose address is never taken),
sub-areas of the Cminor stack block for the current function
(for local non-scalar variables or local scalar variables whose
address is taken), or globally allocated memory areas (for global
variables).
\end{enumerate}
In the first version of the front-end, developed by Zaynah Dargaye and
the authors and published in \cite{2006-Leroy-Blazy-Dargaye}, the
three transformations above were performed in a single pass, resulting
in a large and rather complex proof of semantic preservation.
To make the proofs more manageable, we split the front-end in two
passes: the first performs transformations (1) and (2) above, and the
second performs transformation (3).  A new intermediate language
called \Csharpminor{} was introduced to connect the two passes.
\Csharpminor{} is similar to Cminor, except that it supports a {\tt \&}
operator to take the address of a local variable.  Accordingly, the
semantics of \Csharpminor{}, like that of \Clight{}, allocates one
memory block for each local variable at function entrance, while the
semantics of Cminor allocates only one block.  

To account for this difference in allocation patterns, the proof of
semantic preservation for transformation~(3) exploits the technique of
{\em memory injections} formalized in \cite[section
5.4]{2008-Leroy-Blazy-memory-model}.  It also involves nontrivial
reasoning about separation between memory blocks and between sub-areas
of a block.  The proof requires about 2200 lines of Coq, plus 800
lines for the formalization of memory injections.

The proof of transformations~(1) and~(2) is more routine: since the
memory states match exactly between the original \Clight{} and
the generated \Csharpminor{}, no clever reasoning over memory states,
blocks and pointers is required.  The Coq proof remains relatively
large (2300 lines), but mostly because many cases need to be
considered, especially when resolving overloaded operators.

\section{Validating the \Clight{} semantics} \label{sec:validation}

Developing a formal semantics for a real-world programming language is
no small task; but making sure that the semantics captures the
intended behaviors of programs (as described, for example, by
ISO standards) is even more difficult.  The smallest mistake or
omission in the rules of the semantics can render it incomplete or
downright incorrect.  Below, we list a number of approaches that we
considered to validate a formal semantics such as that of \Clight{}.
Many of these approaches were prototyped but not carried to
completion, and should be considered as work in progress.

\subsection{Manual reviews}

The standard way to build confidence in a formal specification is to
have it reviewed by domain experts.  The size of the semantics for
\Clight{} makes this approach tedious but not downright impossible:
about 800 lines of Coq for the core semantics, plus 1000 lines of Coq
for dependencies such as the formalizations of machine integers,
floating-point numbers, and the memory model.  The fact that the
semantics is written in a formal language such as Coq instead of
ordinary mathematics is a mixed blessing.  On the one
hand, the type-checking performed by Coq guarantees the absence of
type errors and undefined predicates in the specification, while such
trivial errors are common in hand-written semantics.  On the other
hand, domain experts might not be familiar with the formal language
used and could prefer more conventional presentations as \eg inference
rules.  (We have not yet found any C language expert who is
comfortable with Coq, while several of them are fluent with inference
rules.)  Manual transliteration of Coq specifications into \LaTeX{}
inference rules (as we did in this paper) is always possible but can
introduce or (worse) mask errors.  Better approaches include automatic
generation of \LaTeX{} from formal specifications, like Isabelle/HOL
and Ott do \cite{Isabelle-HOL-book,Sewell-Ott-07}.

\subsection{Proving properties of the semantics}

The primary use of formal semantics is to prove properties of programs
and meta-properties of the semantics.  Such proofs, especially when
conducted on machine, are effective at revealing errors in the
semantics.  For example, in the case of strongly-typed languages, type
soundness proofs (showing that well-typed programs do not go wrong)
are often used for this purpose.  In the case of \Clight{}, a type
soundness proof is not very informative, since the type system of C is
coarse and unsound to begin with: the best we could hope for is a
subject reduction property, but the progress property does not hold.
Less ambitious sanity checks include ``common sense'' properties such
as those of the {\tt field{\char95}offset} function mentioned at end of
section~\ref{sec:types}, as well as determinism of evaluation, which
we obtained as a corollary of the verification of the CompCert
compiler \cite[sections 2.1 and 13.3]{2008-Leroy-backend}.

\subsection{Verified translations}

Extending the previous approach to proving properties involving two
formal semantics instead of one, we found that proving semantics
preservation for a translation from one language to another is
effective at exposing errors not only in the translation algorithm,
but also in the semantics of the two languages involved.  If the
translation ``looks right'' to compiler experts and the semantics of
the target language has already been debugged, such a proof of
semantic preservation therefore generates confidence in the semantics
of the source language.  In the case of CompCert, the semantics of the
Cminor intermediate language is smaller (300 lines) and much simpler
than that of \Clight{}; subsequent intermediate languages in the
back-end such as RTL are even simpler, culminating in the semantics of
the PPC assembly language, which is a large but conceptually trivial
transition function \cite{2008-Leroy-backend}.  The existence of
semantic-preserving translations between these languages therefore
constitutes an indirect validation of their semantics.

Semantic preservation proofs and type soundness proofs detect
different kinds of errors in semantics.  For a trivial example, assume
that the \Clight{} semantics erroneously interprets the {\tt +} operator
at type {\tt int} as integer subtraction.  This error would not invalidate
an hypothetical type soundness proof, but would show up immediately in
the proof of semantic preservation for the CompCert front-end,
assuming of course that we did not commit the same error in the
translations nor in the semantics of Cminor.  On the other hand, a
type soundness proof can reveal that an evaluation rule is missing
(this shows up as failures of the progress property). A semantic
preservation proof can point out a missing rule in the semantics of
the target language but not in the semantics of the source language,
since it takes as hypothesis that the source program does not go wrong.

\subsection{Testing executable semantics} \label{s:executable-semantics}

Just like programs, formal specifications can be tested against test
suites that exemplifies expected behaviors.  An impressive example of
this approach is the HOL specification of the
TCP/IP protocol by Sewell \etal \cite{Sewell-TCP-06}, which was
extensively validated against network traces generated by actual
implementations of the protocol.

In the case of formal semantics, testing requires that the semantics
is {\em executable\/}: there must exist an effective way to determine
the result of a given program in a given initial environment.  The Coq
proof assistant does not provide efficient ways to execute a
specification written using inductive predicates such as our semantics
for \Clight{}. (But see \cite{Delahaye-07} for ongoing work in this
direction.)
As discussed in \cite{Appel-Leroy-listmachine-lfmtp}, the
{\tt eauto} tactic of Coq, which performs Prolog-style resolution, can
sometimes be used as the poor man's logic interpreter to execute
inductive predicates.  However, the \Clight{} semantics is too large
and not syntax-directed enough to render this approach effective.

On the other hand, Coq provides excellent facilities for executing
specifications written as recursive functions: an interpreter is built
in the Coq type-checker to perform conversion tests; Coq 8.0 introduced
a bytecode compiler to a virtual machine, speeding up the evaluation
of Coq terms by one order of magnitude \cite{Gregoire-Leroy-02};
finally, the extraction facility of Coq can also be used to generate
executable Caml code.  The recommended approach to execute a Coq
specification by inductive predicates, therefore, is to define a
reference interpreter as a Coq function, prove its equivalence with
the inductive specification, and evaluate applications of the
function.  Since Coq demands that all recursive functions terminate,
these interpretation functions are often parameterized by a
nonnegative integer counter $n$ bounding the depth of the evaluation.
Taking the execution of \Clight{} statements as an example,
the corresponding interpretation function is of the shape
$$ {\tt exec{\char95}stmt}(W, n, G, E, M, s) = 
   {\tt Bottom}(t) \alt {\tt Result}(t, \out, M') \alt {\tt Error} $$
where $n$ is the maximal recursion depth, $G, E, M$ are the initial
state, and $s$ the statement to execute.  The result of execution is
either {\tt Error}, meaning that execution goes wrong, or 
${\tt Result}(t, \out, M')$, meaning that execution terminates with trace $t$,
outcome $\out$ and final memory state $M'$, or ${\tt Bottom}(t)$, meaning
that the maximal recursion depth was exceeded after producing the
partial trace $t$.   To handle the non-determinism introduced by input/output
operations, {\tt exec{\char95}stmt} is parameterized over a {\em world} $W$:
a partial function that determines the result of an input/output operation as a
function of its arguments and the input/output operation previously performed
\cite[section 13.1]{2008-Leroy-backend}.  

The following two properties characterize the
correctness of the {\tt exec{\char95}stmt} function with respect to the inductive
specification of the semantics:
\begin{eqnarray*}
G, E |- \sexecstmt{s}{M}{\out}{t}{M'} ~\wedge~ W \sem t
& \Leftrightarrow &
\exists n,~ {\tt exec{\char95}stmt}(W,n,G,E,M,s) = {\tt Result}(t, \out, M) \\
G, E |- \sexecinfstmt{s}{M}{T} ~\wedge~ W \sem T 
& \Leftrightarrow &
\forall n, \exists t,
\begin{array}[t]{l}
{\tt exec{\char95}stmt}(W,n,G,E,M,s) = {\tt Bottom}(t) \\
\wedge ~ t \mbox{ is a prefix of } T
\end{array}
\end{eqnarray*}
Here, $W \sem t$  means that the trace $t$ is consistent with the
world $W$, in the sense of \cite[section 13.1]{2008-Leroy-backend}.
See \cite{2007-Leroy-Grall} for detailed proofs of these properties
in the simpler case of call-by-value $\lambda$-calculus without
traces.  The proof of the second property requires classical reasoning
with the axiom of excluded middle.

We are currently implementing the approach outlined above, although it
is not finished at the time of this writing.  Given the availability
of the CompCert verified compiler, one may wonder what is gained by
using a reference \Clight{} interpreter to run tests, instead of just
compiling them with CompCert and executing the generated PowerPC
assembly code.  We believe that nothing is gained for test programs
with well-defined semantics.  However, the reference interpreter
enables us to check that programs with undefined semantics do go
wrong, while the CompCert compiler can (and often does) turn them into
correct PowerPC code.

\subsection{Equivalence with alternate semantics}

Yet another way to validate a formal semantics is to write several
alternate semantics for the same language, using different styles of
semantics, and prove logical implications between them.  In the case
of the \cminor{} intermediate language and with the help of Andrew
Appel, we developed three semantics:
(1) a big-step operational semantics in the style of the \Clight{}
semantics described in the present paper \cite{2006-Leroy-compcert}; 
(2) a small-step, continuation-based semantics
\cite{2007-Blazy-Appel,2008-Leroy-backend};
(3) an axiomatic semantics based on separation logic
\cite{2007-Blazy-Appel}.  Semantics (1) and (3) were proved correct
against semantics (2).
Likewise, for \Clight{} and with the help of Keiko Nakata, we
prototyped (but did not complete yet) three alternate semantics to the
big-step operational semantics presented in this paper: (1) a
small-step, continuation-based semantics; (2) the reference
interpreter outlined above; (3) an axiomatic semantics.

Proving the correctness of a semantics with respect to another is an
effective way to find mistakes in both.  For instance, the correctness
of an axiomatic semantics against a big-step operational
semantics without traces can be stated as follows: if $\{P\} s \{Q\}$
is a valid Hoare triple, then for all initial states $G, E, M$
satisfying the precondition $P$, either the statement $s$ diverges
($G, E |- s, M \Rightarrow \infty$) or it terminates ($G, E |- s, M
\Rightarrow \out, M'$) and the outcome $\out$ and the final state $G,
E, M'$ satisfy postcondition $Q$.  The proof of this property
exercises all cases of the big-step operational semantics and is
effective at pointing out mistakes and omissions in the latter.
Extending this approach to traces raises delicate issues that we have
not solved yet.  First, the axiomatic semantics must be extended with
ways for the postconditions $Q$ to assert properties of the traces
generated by the execution of the statement $s$.  A possible source of
inspiration is the recent work by Hoare and O'Hearn
\cite{Hoare-OHearn-08}.  Second, in the case of a loop such as
$\{P\} ~ {\tt while}(a)~s ~ \{Q\}$, we must not only show that the loop either
terminates or diverges without going wrong, as in the earlier
proof, but also prove the existence of the corresponding traces of
events.  In the diverging case, this runs into technical problems
with Coq's guardedness restrictions on coinductive
definitions and proofs.

In the examples given above, the various semantics were written by the
same team and share some elements, such as the memory model and the
semantics of \Clight{} expressions.  Mistakes in the shared parts will
obviously not show up during the equivalence proofs.  Relating two
independently-written semantics would provide a more convincing
validation.  In our case, an obvious candidate for comparison with
\Clight{} is the Cholera semantics of Norrish \cite{Norrish:phd}.
There are notable differences between our semantics and Cholera,
discussed in section~\ref{sec:sota}, but we believe that our semantics
is a refinement of the Cholera model.  A practical issue with
formalizing this intuition is that Cholera is formalized in HOL while
our semantics is formalized in Coq.

\section{Related work}\label{sec:sota}

\paragraph{Mechanized semantics for C}

The work closest to ours is Norrish's Cholera project
\cite{Norrish:phd}, which formalizes the static and dynamic semantics
of a large subset of C using the HOL proof assistant.  Unlike
\Clight{}, Cholera supports side effects within expressions and
accounts for the partially specified evaluation order of~C.
For this purpose, the semantics of expressions is given in small-step
style as a non-deterministic reduction relation, while the semantics
of statements is given in big-step style.  Norrish used this semantics
to characterize precisely the amount of non-determinism allowed by the
C~standard \cite{Norrish-99}.  Also, the memory model underlying
Cholera is more abstract than that of \Clight{}, leaving unspecified a
number of behaviors that \Clight{} specifies.

Tews {\em et al} \cite{Tews-08,Robin-D13} developed a denotational
semantics for a subset of the \Cplusplus\ language.
The semantics is presented as a shallow embedding in the PVS
prover.  Expressions and statements are modeled as state transformers:
functions from initial states to final states plus value (for
expressions) or outcome (for statements).  The subset of
\Cplusplus\ handled is close to our Clight, with a few differences:
side effects within expressions are allowed (and treated using a fixed
evaluation order); the behavior of arithmetic operations in case of
overflow is not specified; the {\tt goto} statement is not handled, but
the state transformer approach could be extended to do so \cite{Tews-05}.

Using the Coq proof assistant, Gim\'enez and Ledinot~\cite{C:Dassault} define 
a denotational semantics for a subset of~C appropriate as target
language for the compilation of the Lustre synchronous dataflow language. 
Owing to the particular shape of Lustre programs, the subset of~C does
not contain general loops nor recursive functions, but only counted
{\tt for} loops.  Pointer arithmetic is not supported.  

As part of the Verisoft project~\cite{Verisoft-project},
the semantics of a subset of C called C0 has been formalized using
Isabelle/HOL, as well as the correctness of a compiler from C0 to DLX
assembly language \cite{paul:sefm,strecker05:_compil_verif_c0,Schirmer-PhD}.
C0 is a type-safe subset of C, close to Pascal, and significantly smaller than
\Clight{}: for instance, there is no pointer arithmetic, nor {\tt break}
and {\tt continue} statements.  A big-step semantics and a small-step
semantics have been defined for C0, the latter enabling reasoning about 
non-terminating executions. 

\paragraph{Paper and pencil semantics for C}

Papaspyrou~\cite{papaspyrou98} develops a monadic denotational
semantics for most of ISO~C.  Non-determinism in expression evaluation
is modeled precisely.  The semantics was validated by testing with the
help of a reference interpreter written in Haskell.

Nepomniaschy \etal~\cite{2003-Nepomniaschy} define a big-step semantics for a 
subset of C similar to Pascal: it supports limited uses of {\tt goto}
statements, but not pointer arithmetic.

Abstract state machines have been used to give semantics for C~\cite{gurevich:C} 
and for C{\tt \#}~\cite{borger:C}. The latter formalization is arguably the most
complete (in terms of the number of language features handled) formal
semantics for an imperative language.

\paragraph{Other examples of mechanized semantics}

Proof assistants were used to mechanize semantics for languages that
are higher-level than~C.  Representative examples include
\cite{InwegenG93,LeeCH07} for Standard ML, \cite{Owens08} for a subset
of OCaml, and \cite{KleinN06} for a subset of Java.  Other
Java-related mechanized verifications are surveyed in \cite{Hartel-Moreau-01}.
Many of these semantics were validated by conducting type soundness
proofs.  

\paragraph{Subsets of C}

Many uses of C in embedded or critical applications mandate strict
coding guidelines restricting programmers to a ``safer'' subset of C
\cite{Hatton-04}.  A well-known example is MISRA~C \cite{Misra-C}.
MISRA~C and \Clight{} share some restrictions (such as
structured {\tt switch} statements with {\tt default} cases at the end), but 
otherwise differ significantly.  For instance, MISRA~C prohibits
recursive functions, but permits all uses of {\tt goto}.  More
generally, the restrictions of MISRA~C and related guidelines are
driven by software engineering considerations and the desire for
tool-assisted checking, while the restrictions of \Clight{} stem from
the desire to keep its formal semantics manageable.

Several tools for static analysis and deductive verification of C
programs use simplified subsets of C as intermediate representations.
We already discussed the CIL intermediate representation \cite{cil}.
Other examples include the Frama-C intermediate representation
\cite{framac}, which extends CIL's with logical assertions,
and the Newspeak representation
\cite{2008-Hymans-Levillain}.  CIL is richer than \Clight{} and
accurately represents all of ISO~C plus some extensions.  Newspeak is
lower-level than \Clight{} and targeted more towards
static analysis than towards compilation.

\section{Conclusions and future work}\label{sec:concl}

In this article, we have formally defined the \Clight{} subset of the C
programming language and its dynamic semantics.  While there is
no general agreement on the formal semantics of the C language,
we believe that \Clight{} is a reasonable proposal that works well in
the context of the formal verification of a compiler.  We hope 
that, in the future, \Clight{} might be useful in other contexts
such as static analyzers and program provers and their formal
verification.

Several extensions of \Clight{} can be considered.  One direction,
discussed in \cite{2008-Leroy-Blazy-memory-model}, is to relax
the memory model so as to model byte- and bit-level accesses to in-memory
data representations, as is commonly done in systems programming.

Another direction is to add support for some of the C constructs
currently missing, in particular the {\tt goto} statement.  The main issue
here is to formalize the dynamic semantics of {\tt goto} in a way that
lends itself well to proofs.  Natural semantics based on statement
outcomes can be extended with support for {\tt goto} by following the
approach proposed by Tews \cite{Tews-05}, but at the cost of nearly
doubling the size of the semantics.  Support for {\tt goto} statements is
much easier to add to transition semantics based on continuations, as
the Cminor semantics exemplifies \cite[section~4]{2008-Leroy-backend}.
However, such transition semantics do not lend themselves easily to
proving transformations of loops such as those performed by the
front-end of \compcert{} (transformation~2 in
section~\ref{s:compiling-clight}).

Finally, the restriction that \Clight{} expressions are pure is both a
blessing and a curse: on the one hand, it greatly simplifies all further
processing of \Clight{}, be it compilation, static analysis or program
verification; on the other hand, programmers cannot be expected to
directly write programs where all expressions are pure, requiring
nontrivial, untrusted program transformations in the \Clight{} parser.
One way to address this issue would be to define an extension of
\Clight{}, tentatively called Cmedium, that supports side effects within
expressions, and develop and prove correct a translation
from Cmedium to \Clight{}.

\bibliographystyle{spmpsci}

\end{document}